\documentclass{aa}
\usepackage{psfig}
\usepackage{latexsym}
\usepackage{graphicx}

\def\bron{GX~17+2}
\def\ecs{erg~cm$^{-2}$s$^{-1}$}
\def\lum{erg~s$^{-1}$}
\renewcommand{\tabcolsep}{1.5mm} 

\begin{document}

\title{Superbursts at near-Eddington mass accretion rates}

\titlerunning{Superbursts at near-Eddington mass accretion rates} 
\authorrunning{J.J.M. in 't Zand, R. Cornelisse \& A. Cumming}

\author{
J.J.M.~in~'t~Zand\inst{1,2},
R.~Cornelisse\inst{3} \&
A. Cumming\inst{4}\thanks{Hubble Fellow}
}

\offprints{J.J.M. in 't Zand, email {\tt jeanz@sron.nl}}

\institute{     SRON National Institute for Space Research, Sorbonnelaan 2,
                NL - 3584 CA Utrecht, the Netherlands 
	 \and
                Astronomical Institute, Utrecht University, P.O. Box 80000,
                NL - 3508 TA Utrecht, the Netherlands
         \and
                Dept. of Physics and Astronomy, University of Southampton,
                Hampshire SO17 1BJ, U.K.
         \and
                Dept. of Astronomy and Astrophysics, University of
                California, Santa Cruz, CA 95064, U.S.A.
	}

\date{Accepted July 2004}

\abstract{Models for superbursts from neutron stars involving carbon
shell flashes predict that the mass accretion rate should be anywhere
in excess of one tenth of the Eddington limit. Yet, superbursts have
so far only been detected in systems for which the accretion rate is
limited between 0.1 and 0.25 times that limit. The question arises
whether this is a selection effect or an intrinsic property.
Therefore, we have undertaken a systematic study of data from the
BeppoSAX Wide Field Cameras on the luminous source \bron, comprising
10 Msec of effective observing time on superbursts. \bron\ contains a
neutron star with regular Type-I X-ray bursts and accretes matter
within a few tens of percents of the Eddington limit. We find four
hours-long flares which reasonably match superburst
characteristics. Two show a sudden rise (i.e., faster than 10~s), and
two show a smooth decay combined with spectral softening. The implied
superburst recurrence time, carbon ignition column and quenching time
for ordinary bursts are close to the predicted values. However, the
flare decay time, fluence and the implied energy production of
(2--4)$\times10^{17}$~erg~g$^{-1}$ are larger than expected from
current theory.  }

\maketitle 

\section{Introduction}
\label{intro}

Superbursts are X-ray flares with a rise of a few seconds and an
exponential-like decay of a few hours. They originate from neutron
stars that accrete matter from a low-mass companion star. Thus far,
nine superbursts have been detected from seven sources: one from
4U~1735-44 (Cornelisse et al. 2000), Ser X-1 (Cornelisse et al. 2002),
KS~1731-260 (Kuulkers et al. 2002a), 4U~1820-303 (Strohmayer \& Brown
2002), GX 3+1 (Kuulkers 2002) and 4U~1254-690 (In~'t~Zand et
al. 2003a), and three from 4U~1636-536 within 4.7~yr (Wijnands 2001;
Strohmayer \& Markwardt 2002; Kuulkers et al. 2004). The long duration
sets them apart from 'ordinary' Type-I X-ray bursts which are 10$^3$
times shorter, less fluent and more frequent. The joint characteristic
of the superburst sources is that they also exhibit ordinary Type-I
X-ray bursts, show evidence for stable helium burning (In~'t~Zand et
al. 2003a) and accrete at levels between 0.1 and 0.25 times the
Eddington limit. For a recent review, see Kuulkers (2004).

\begin{figure*}[t]
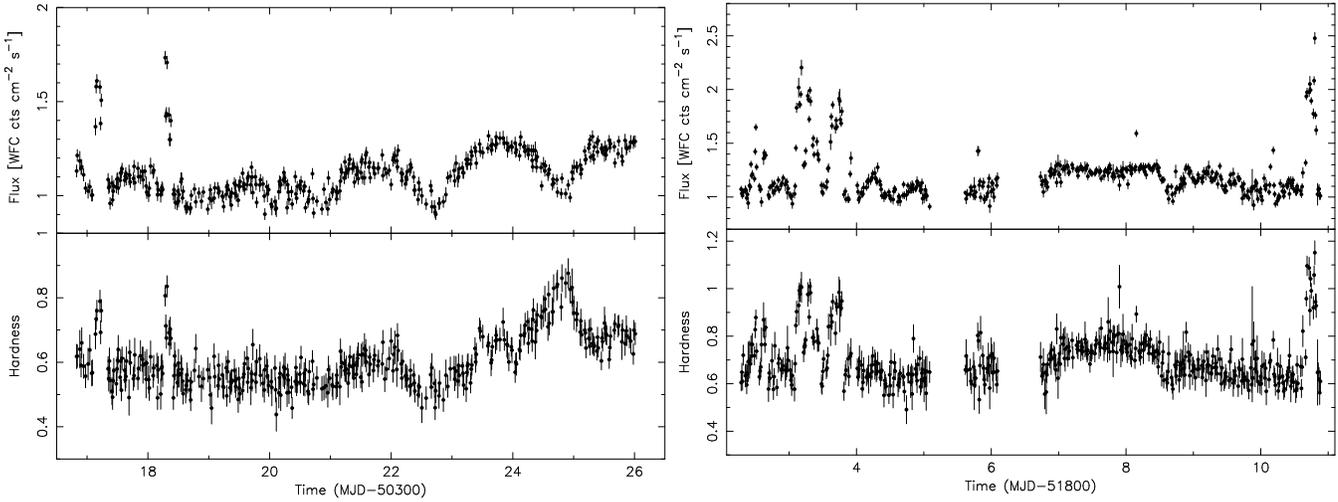

\includegraphics[height=\columnwidth,angle=270]{figlc8XX.ps}
\includegraphics[height=\columnwidth,angle=270]{figlc9XXX.ps}
\caption[]{Flux and 2-6/6-25 keV spectral hardness ratios of \bron\
during two nine-day long observations in August 1996 (left) and
September 1999 (right). WFC unit 2 was employed in both
observations. The time resolution is 15 min.
\label{figlcs}}
\end{figure*}

\begin{table}[!b]
\caption[]{List of 25 ordinary Type-I X-ray bursts detected from
\bron\ during the WFC monitoring campaign. 13 were detected with the
WFCs, 8 with RXTE/PCA (Kuulkers et al. 2002b), 2 with BeppoSAX/NFI (Di
Salvo et al. 2000) and 2 with both PCA and NFI. The e-folding decay
times were determined from observed photon rates in the complete
bandpass of each instrument (PCA for the 2 PCA\&NFI bursts) and serve
purely as a guide.\label{tab0}}
\begin{center}
\renewcommand{\tabcolsep}{0.5cm}
\begin{tabular}{lcl}
\hline
Time & Instrument & Decay time (s) \\
\hline
50121.67861 & PCA  & $1.83\pm0.08$\\
50342.48699 & WFC  & $550\pm100$ \\
50343.79428 & WFC  & $100\pm20$ \\
50383.64851 & WFC  & $90\pm20$ \\
50487.10873 & PCA  & $248^4_{-9}$ \\
50538.48909 & WFC  & $100\pm40$ \\
50552.15754 & WFC  & $400\pm100$ \\
50716.15211 & WFC  & $400\pm100$ \\
51032.55267 & PCA  & $2.55\pm0.24$ \\
51084.51257 & WFC  & $20\pm10$ \\
51135.36905 & PCA  & $197\pm2$ \\
51135.60934 & PCA  & $2.06\pm0.13$ \\
51454.65037 & PCA  & $274\pm3$ \\
51456.98730 & PCA \&NFI & $77.3\pm1.2$ \\
51457.46566 & PCA \&NFI & $70.2\pm1.4$ \\
51457.99918 & NFI & $124\pm15$ \\
51460.01504 & NFI & $140\pm60$ \\
51460.52389 & PCA  & $76.4\pm1.5$ \\
51461.38249 & PCA  & $115\pm3$ \\
51618.92595 & WFC  & $240\pm120$ \\
51808.14798 & WFC  & $300\pm100$ \\
51989.72422 & WFC  & $97\pm52$ \\
52003.73772 & WFC  & $226\pm115$ \\
52004.15411 & WFC  & $60\pm20$ \\
52172.68849 & WFC  & $150\pm75$ \\
\hline
\end{tabular}
\end{center}
\end{table}

Superbursts are attributed to unstable carbon shell burning (Woosley
\& Taam 1976; Strohmayer \& Brown 2002) within the top 100~m of the
neutron star, as compared to unstable helium/hydrogen burning within
the top 10~m for ordinary bursts (Hansen \& van Horn 1975; Maraschi \&
Cavaliere 1977; Joss 1977; Taam \& Picklum 1979; Fujimoto et
al. 1981).  They are an important diagnostic because they allow us to
probe the products of rp-process H/He burning. Except perhaps for
4U~1820-303 which likely harbors a hydrogen-deficient dwarf (e.g.,
Cumming 2003), the carbon is thought to be located in a heavy-element
ocean that is produced by the rp-process during hydrogen burning in
higher-up layers (Cumming \& Bildsten 2001; CB01). The energy released
during a superburst and the decay time (Cumming \& Macbeth 2004; CM04)
are well explained by this model.  Indeed, the superburst duration is
naturally understood as the thermal timescale of a layer 1000 times
thicker than a usual Type I X-ray burst, as required to satisfy the
observed superburst energies.  Uncertainties exist about the ignition,
the fuel production and the mixture. Possibly,
photo-disintegration-triggered nuclear energy release plays an
important role (Schatz et al. 2003a, but see Woosley et al. 2004).

Cumming \& Bildsten (2001) predict that for {\em any} luminosity above
0.1 times Eddington superbursts should occur. Furthermore, at higher
persistent levels the superburst rate would increase and the burst
fluence decrease.  As was already pointed out by CB01, it is more
difficult to find superbursts in systems that radiate near the
Eddington limit than in lower luminosity systems, because the dynamic
range is much smaller. If the system is emitting at 90\% of the
Eddington limit, the signal-to-background ratio is 0.1 at maximum
while it may be as high as 10 if the persistent luminosity level is
near 10\% of the Eddington limit. Furthermore, the amplitude of the
hour-time-scale variability in the more luminous systems may be
comparable to the peak flux of superbursts.

The high-luminosity system that is probably best suited to search for
superbursts and test the predictions by CB01 is
\bron. \bron\ is one of two Galactic Z sources that exhibit X-ray
bursts, the other being Cyg X-2. Z sources trace a Z-shaped path in
X-ray color-color and color-intensity diagrams (e.g., Hasinger et
al. 1990). It is believed that they radiate close (i.e., within a few
tens of percents) to the Eddington limit.  Recently, two detailed
X-ray studies have been performed of \bron\ with a large volume of
data taken with the Proportional Counter Array (PCA) on the Rossi
X-ray Timing Explorer (RXTE). Homan et al. (2002) and Kuulkers et al.
(2002b) find that 1) \bron\ spends 28\% of the time on the flaring
branch of the Z- diagram where it is thought that the luminosity
exceeds the Eddington limit; 2) all 10 detected X-ray bursts occur on
the normal branch in which the source spends 44\% of the time,
implying an average burst recurrence time of 8 hours; 3) 5 X-ray bursts
exhibit photospheric radius-expansion due to fluxes near the Eddington
limit implying a distance between 12 kpc (for isotropic burst emission)
and 8 kpc (for anisotropic emission).
No superbursts have been reported
from these data. The companion star to the neutron star in \bron\ has
not been identified. An infrared counterpart was detected but the
radiation is unrelated to the companion star (Callanan et al. 2002).

\renewcommand{\tabcolsep}{0.15cm}

\begin{figure*}[t]
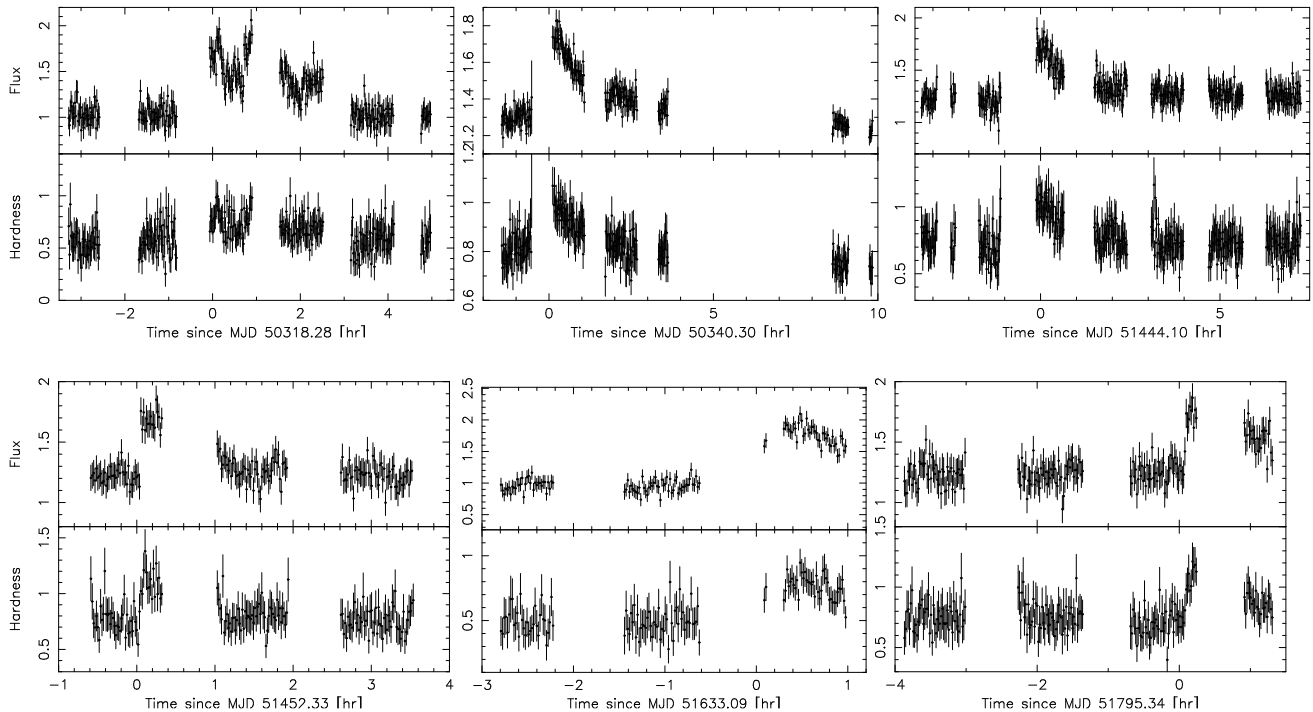

\includegraphics[width=0.5\columnwidth,angle=270.]{llc1.ps}
\includegraphics[width=0.5\columnwidth,angle=270.]{llc2.ps}
\includegraphics[width=0.5\columnwidth,angle=270.]{llc3.ps}

\vspace{0.5cm}
\includegraphics[width=0.5\columnwidth,angle=270.]{llc4.ps}
\includegraphics[width=0.5\columnwidth,angle=270.]{llc5.ps}
\includegraphics[width=0.5\columnwidth,angle=270.]{llc6.ps}
\caption{2-28 keV flux and 2-6/6-25 keV hardness histories of all 6
isolated flares from \bron\ with 64~s time resolution. All data of
each flare are plotted (which is somewhat limited in the last two
flares).\label{figsblc}}
\end{figure*}

We here report on a superburst search in data from the BeppoSAX Wide
Field Cameras. The likelihood of success was anticipated to be
much higher than for the RXTE data, because of a nine times longer net
exposure time.

\section{Observations and characterization of variability}
\label{data}

The X-ray observatory BeppoSAX (Boella et al. 1997) operated from June
1996 to April 2002 and carried two identical Wide Field Cameras (WFCs;
Jager et al. 1997) with $40^{\rm o}\times40^{\rm o}$ fields of view
and 5\arcmin\ angular resolution in a 2-25 keV bandpass.  The WFCs
observed \bron\ during twelve semi-yearly campaigns of the Galactic
center region (e.g., In 't Zand et al. 2004a) with a total net
exposure time of 5.6 Msec. The sum of the elapse time of all
observations is 11.3~Msec. This includes the data gaps due to earth
occultations and passages through the South Atlantic Geomagnetic
Anomaly which are so short that they do not degrade the detection
capability for superbursts. During most observations the source was at
a relatively large off-axis angle of about 16\degr. The implied
reduction in sensitivity has only a limited detrimental effect on the
detection capability because the source is bright; it can be detected
in exposures as short as 10~s.

In Fig.~\ref{figlcs} we show time histories of the flux and spectral
hardness of \bron\ during two exceptionally long observations lasting
nine days. These nicely illustrate the typical source behavior
inherent to a Z source, namely an inverse hardness versus flux
relationship (e.g., at MJD 50324) when the source is in the
'horizontal branch' of the Z pattern, a proportional hardness versus
flux relationship in the 'normal branch', and flaring on a time scale
of hours in the 'flaring branch'. This branch behavior in \bron\ has
been studied in detail by Homan et al. (2002).

Furthermore, 13 ordinary X-ray bursts were detected. These have
e-folding decay times of a few hundred seconds. The WFC data are not
of high-enough quality to detect short X-ray bursts (i.e., shorter
than 10~s and peak fluxes of the order of 0.5 Crab) like the ones
observed by Kuulkers et al. (2002b) in 30\% of all cases. Table~\ref{tab0}
lists all 25 ordinary bursts that were detected during the period that
the WFCs monitored \bron.

\section{Search for superbursts}
\label{sb}

\begin{table*}
\begin{center}
\caption[]{Isolated flares from \bron\ as seen with the
WFCs.\label{tab1} The arrows indicate the trend in k$T$ from flare
start to end.}

\begin{tabular}{lllllllllll}
\hline\hline
No. & Time & Obs. & Rise & Obs. photon peak & Decay       & $kT_{\rm peak}$& $R_{\rm 8~kpc}$  & Peak       & Fluence$^\ddag$\\
 &     & period & time  & flux (WFC& time     &    (keV)              & (km)                  & flux$^\dag$&           \\
 &     &    & (s)      & c~s$^{-1}$cm$^{-2}$)&(hr) &                  &                       &            &                \\
\hline
1 & 50318.28 &  816 & $<2900$ & $0.95\pm0.10$ & $1.5\pm0.2$ & $2.0\pm0.1\rightarrow2.1\pm0.2$ & 5.5 & 1.6 &  \\
2 & 50340.30 &  927 & $<2300$ & $0.45\pm0.05$ & $1.9\pm0.1$ & $2.2\pm0.1\rightarrow1.7\pm0.1$ & 5.5 & 0.8 & $5.5\pm1.3$ \\
3 & 51444.10 & 7615 & $<2300$ & $0.52\pm0.04$ & $1.0\pm0.2$ & $2.4\pm0.1\rightarrow1.5\pm0.3$ & 4.5 & 1.1 & $4.0\pm0.6$ \\
4 & 51452.33 & 7673 & $\la 10$   & $0.56\pm0.02$ & $0.7\pm0.1$ & $3.1\pm0.3\rightarrow3.0\pm0.8$ & 1.9  & 1.4 & $4.4\pm1.0$ \\
5 & 51633.09 & 8813 & $<2600$ & $1.00\pm0.05$ & $0.9\pm0.3$ & $1.9\pm0.1$                    & 7.3  & 1.7  \\
6 & 51795.34 & 9764 & $\la 10$   & $0.54\pm0.10$ & $2.2\pm0.6$ & $2.4\pm0.5\rightarrow1.9\pm0.2$ & 5.0  & 1.5 & $>4.2\pm1.0$ \\
\hline\hline
\end{tabular}
\end{center}
$^\dag$in $10^{-8}$~\ecs; $^\ddag$in $10^{-5}$~erg~cm$^{-2}$. The
fluence is established by adding the bolometric unabsorbed fluences
per time interval over which a spectrum was accumulated, and
interpolating the bolometric fluxes for the data gaps. For data gaps
that occur immediately before the observed start of a flare, it was
assumed that the flare starts at mid time; an uncertainty is added to
the total which represents the fluence from mid time to observed
onset. The fluence of the last flare is formally unconstrained because
of a premature end of observations, but it is expected not to exceed
12$\times10^{-5}$erg~cm$^{-2}$ if the flare decays with the e-folding
decay time as measured from the available data. Fluences were not
determined for flares which are unlikely to be superbursts (flares 1
and 5, see text).
\end{table*}

Superbursts are characterized by a rise which lasts only a few
seconds, a smooth exponential-like decay with an e-folding decay time
between about 1 and 6 hours and a black body spectrum with a peak
temperature of 2.5 to 3 keV which decreases during the decay (e.g.,
review by Kuulkers 2004). Apart from the long decay time these
characteristics match those of ordinary Type-I X-ray bursts (e.g.,
reviews by Lewin et al. 1993 and Strohmayer \& Bildsten 2003).

In order to identify superbursts in \bron\ it is crucial to
discriminate against flares that are thought to result from quick
changes in the mass accretion rate. This is not trivial. Flares have
the same durations and exhibit similar few-keV black-body spectra
which cool during decay (e.g., Hoshi \& Asaoka 1993). The best
discriminator is perhaps the time profile of the intensity:
superbursts always have fast rises (i.e., less than 10~s) while flares
generally do not (rise times usually are several minutes or longer)
and superbursts always have smooth decays without reflaring in
contrast to flares.  Another difference may be in the high-frequency
timing behavior.  In the flaring branch a broad ($>$50\% FWHM)
low-amplitude (2-6\% rms) 14-23 Hz QPO (``FBO'') is present while
low-frequency QPOs are absent during ordinary Type-I X-ray bursts
(Homan et al.  2002) and, therefore, possibly superbursts. However,
the WFC data are of insufficient statistical quality to detect FBOs.

Given these difficulties we considered it best to search for
superburst candidates in data stretches where the source is not in an
obvious continued flaring state. We define such a state as when \bron\
shows variability larger than a factor of two on a time scale of an
hour and for a duration in excess of 6 hours. An example is visible in
the right-hand panel of Fig.~\ref{figlcs}, at MJD~51803. We find that
the flaring state pertains to about 10\% of all data\footnote{This
compares to 28\% in the RXTE data, but that has a nine times smaller
exposure time; the 10\% number is supported by checking the 31,231
dwell measurements gathered by the RXTE All-Sky Monitor in 7.9 years
of monitoring \bron}. Thus, fortunately, disregarding the flaring
state episodes preserves 90\% of the exposure. We searched and
identified in the non-flaring state 6 isolated flares which appeared
to show fast rises and near to monotonic decays. They are tabulated in
Table~\ref{tab1} and plotted in Fig.~\ref{figsblc}.  One of these
occurred during the first nine-day observation shown in
Fig.~\ref{figlcs} (left panel; 2nd flare).

Due to the low-earth BeppoSAX orbit, data gaps are frequent. Combined
with the sometimes short observations, the flare onsets have only been
covered in 2 flares and the decay is only partly covered in 2 cases.
The fourth flare is the only one which is well covered at the onset
and decay although there is a relatively large data gap in this flare
as well. We note that \bron\ was not observed by RXTE/PCA during any
of these 6 flares.

\begin{figure}[t]
\psfig{figure=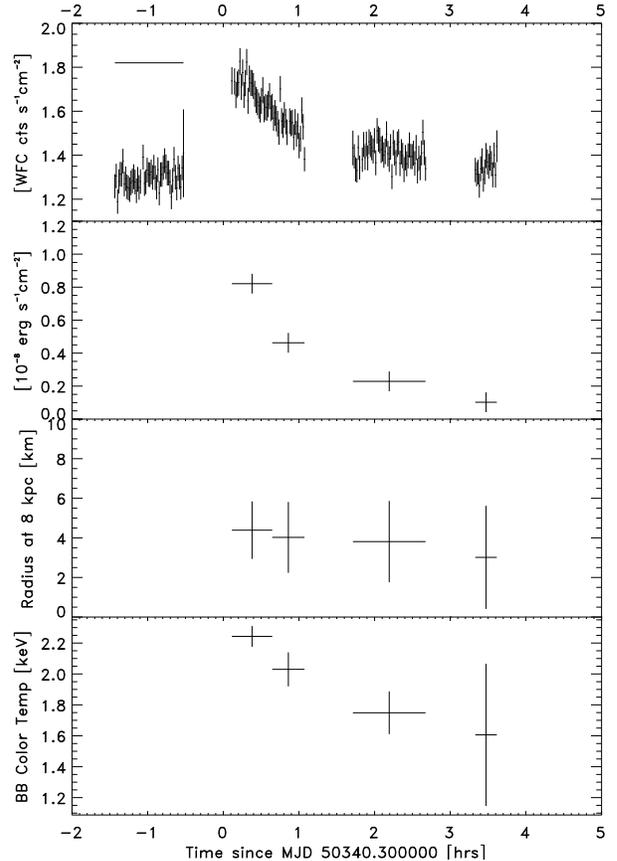,width=\columnwidth,clip=t}
\caption{Time-resolved spectroscopy of the 2nd flare. The horizontal
bar in the top panel indicates the accumulation interval for the 
spectrum of the persistent emission. The time resolution in the top
panel is 64~s, like in Fig.~\ref{figsblc}.\label{figspec927}}
\end{figure}

\section{Assessment of good superburst candidates}

We carried out time-resolved spectroscopy of all six flares.  Time
intervals were defined with statistically meaningful durations during
the flare and one for the final few hours before the flare. The
spectrum was satisfactorily modeled by a combination of a single
(canonical) bremsstrahlung component (suitable to describe the
persistent emission), and a black body radiation component whose
parameters were allowed to vary during the flare and were fixed to
zero during the pre-flare interval. We included interstellar
absorption following Morrison \& McCammon (1983). In Table~\ref{tab1}
we list the trends that we find for the black body temperature and
area (for a distance of 8 kpc) and in Fig.~\ref{figspec927} the full
details are shown for the second flare. The net flare emission could
always be modeled by a few-keV black body.

\begin{figure}[!t]
\includegraphics[height=\columnwidth,angle=270.]{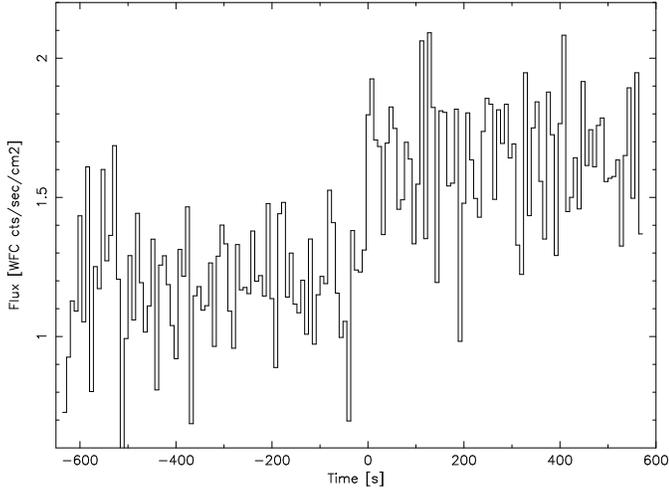}
\caption{Rising part of flare 4 at 8-sec resolution.\label{figrise7673}}
\end{figure}

\begin{figure}[t]
\includegraphics[height=\columnwidth,angle=270.]{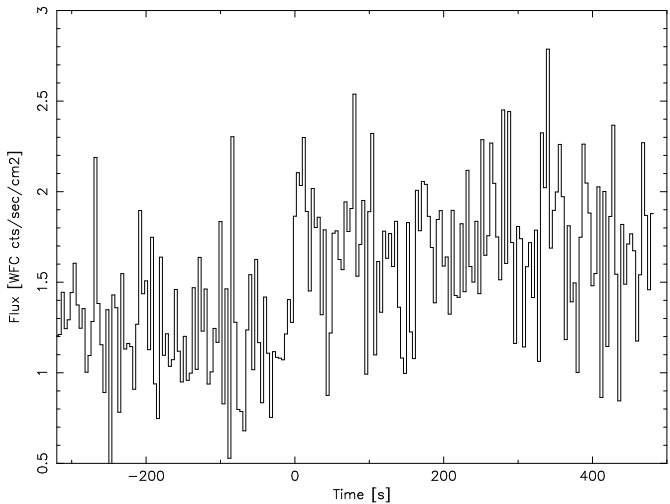}
\caption{Rising part of flare 6 at 4-sec resolution.\label{figrise9764}}
\end{figure}

\begin{figure}[!t]
\includegraphics[width=\columnwidth]{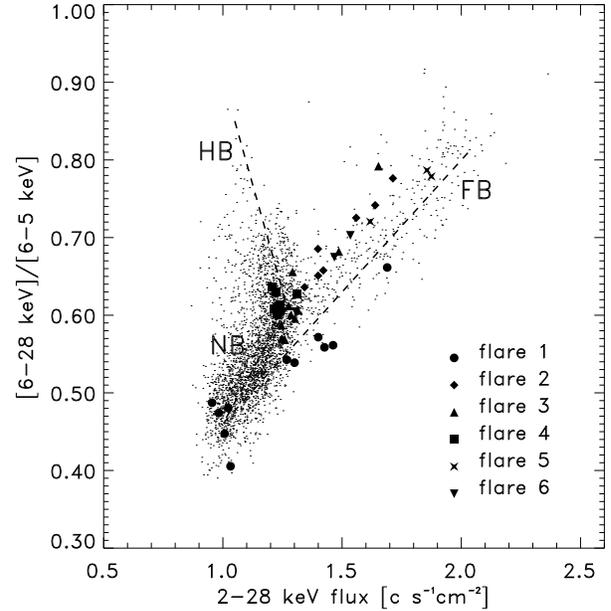}
\caption{Color-intensity diagram as composed from all WFC data. Each
data point represents a 15~min exposure. A good-quality filter has been
applied, eliminating all data points with flux accuracies worse than
0.05~c~s$^{-1}$cm$^{-2}$ and color accuracies worse than 0.05. The 
dashed line indicates the three branches.
\label{figci}}
\end{figure}

Only for two flares (2 and 3) could cooling be proven during the decay
phase.  However, only for the first flare the absence of cooling is
significant. Flares 4 and 6 do show spectral softening during decay
when parameterized with a simple spectral hardness ratio (see
Fig.~\ref{figsblc}). The time profile of the first flare, furthermore,
is not as smooth as one would expect for a superburst.  It clearly
shows upward fluctuations two times in a manner never seen in a
superburst. We disregard the first flare as a superburst candidate.

The flare onset was covered in two cases: flares 4 and 6. Both have a
sharp rise lasting shorter than 10 sec (see Figs.~\ref{figrise7673}
and \ref{figrise9764}). This is consistent with a superburst
identification. The statistical quality is rather limited, but the
high-resolution data also suggest the presence of an ordinary Type-I
X-ray burst of the short variety at the superburst onset. Such bursts
are often referred to as 'precursors' in superburst research, although
they have never been shown to truly precede the superburst. They
reinforce the superburst identification of these two flares.

The one remaining flare for which the rise was not observed and for
which cooling could not be proven is number 5. This flare has one
further characteristic which makes it doubtful as a superburst: like
flare number 1 it has a peak flux twice as large as all other
flares. Therefore, we exclude flare number 5 as a superburst
candidate.

In Fig.~\ref{figci} we show the color-intensity diagram for all data
at a 15~min resolution (same resolution as in Fig.~\ref{figlcs}). The
data points for the six flares are highlighted. The normal and
horizontal branch are visible (cf, Homan et al. 2002), as are
excursions to higher fluxes of up to 2.2~c~s$^{-1}$cm$^{-2}$. There
appear to be two tracks of high-flux excursions, one starting from the
lower part of the normal branch and the other from the vertex between
the normal and horizontal branch. The former track must be the flaring
branch, because flaring branches always start at the lower color end
of the normal branch. The first flare is on this track, and the fifth
flare may also be. The other track is covered with the other four
flares which were identified as good superburst candidates. Thus, this
diagnostic appears to confirm the conclusions regarding the nature of
the 6 flares based on the time profile and black body temperature.

We note that all superbursts appear to spawn from the vertex between
the horizontal and normal branches. This is in contrast to ordinary
bursts which can spawn from anywhere on the normal branch (Kuulkers et
al. 2002b). 

We conclude that of the six investigated flares four are good
superburst candidates: flares 2, 3, 4 and 6.

\section{Discussion}

\subsection{Bolometric luminosity and mass accretion rate}

\bron\ is so interesting for the understanding of superbursts
because it is more luminous than any other superburster and,
therefore, presumably has a higher mass accretion rate. The arguments
for the luminosity being so high are mostly indirect: they are based
on a Z pattern traced in the color-color and color-intensity diagrams
(e.g., Homan et al. 2002) and on the low flux ratio, compared to the
other superbursters, of peaks of ordinary bursts to out-of-burst
episodes (Kuulkers et al. 2002b). The flux ratio is about 0.7 for GX
17+2, while it is at least 1.5 for all other superbursters.

It is quite difficult to infer directly the broad-band luminosity of
GX~17+2 with a better-than-10\% accuracy, as is the case for most
LMXBs in the Galaxy. For an important part this is due to a poorly
constrained distance (8-12 kpc; Kuulkers et al. 2002b). Another source
of uncertainty is the unfamiliarity of the intrinsic spectrum below
2~keV due to the large interstellar absorbing column.  Di Salvo et
al. (2000) made an accurate broadband spectral study of \bron\ on the
basis of a 5-d observation with the BeppoSAX Narrow Field
Instruments. They found that the observed 0.1-200 keV flux was between
1.58$\times10^{-8}$ (at the vertex of the normal and flaring branches)
and 1.84$\times10^{-8}$~\ecs\ (at the vertex of the normal and
horizontal branch), while the source was not caught in the flaring
state. Correcting the fluxes for absorption is difficult because of
the presence of a steep power-law component and the uncertainty about
its low-energy cut off.  Straightforward extrapolation of the power
law to 0.1 keV results in total unabsorbed 0.1-200 keV fluxes of
1.92$\times10^{-8}$ and 3.63$\times10^{-8}$~\ecs\ respectively. If the
power extends to a more realistic 2 keV, the latter value becomes
2.26$\times10^{-8}$~\ecs\ (Di Salvo, priv. comm.). We regard the
latter value as the most reliable unabsorbed maximum flux over the
normal and horizontal branches. For a distance range of 8 to 12 kpc
and assuming isotropic radiation, it translates to a luminosity
between 1.7 and $3.9\times10^{38}$~\lum. The absolute minimum 0.1-200
luminosity is derived from the mimimum observed flux of
1.58$\times10^{-8}$~\ecs\ and the shortest allowable distance of 8
kpc, resulting in $1.2\times10^{38}$~\lum.

The Eddington limit of a neutron star is determined by its size, mass,
and the hydrogen content of its photosphere.  The latter must be high
because some ordinary bursts are relatively long which presumably is
due to prolonged nuclear reactions involving copious free protons
(e.g., Fujimoto et al. 1981). For a canonical neutron star with a mass
of 1.4~$M_\odot$, a radius of 10 km and a solar composition of the
photosphere (i.e., with a hydrogen mass fraction of $X=0.7$), the
Eddington limit is $2\times10^{38}$~\lum\ (see also Sect.
\ref{s53}). Therefore, we conclude that the persistent emission is
with reasonable certainty always larger than 60\% of the Eddington
luminosity and on average perhaps 80--90\%. This is at least three
times higher than for any other superburster.

We note that the net bolometric peak flux as measured by Kuulkers et
al. (2002b) for 5 ordinary Eddington-limited X-ray bursts from \bron\
ranges between 1.4 and $1.7\times10^{-8}$~\ecs\ (as measured on a
0.25~s time scale). Averaged over the complete expansion phases, the
fluxes range between 1.2 and 1.3$\times10^{-8}$~\ecs. These presumed
Eddington-limited values are 40 to 50\% smaller than the
above-mentioned values for the persistent emission (more if the power
extends to below 2 keV). Kuulkers et al. (2002b), on the basis of a
similar comparison with persistent emission measured with RXTE in 3-20
keV, propose that this discrepancy can be explained if the burst
emission is anisotropic (i.e., obscuration of part of the burst
emission) and if the distance is 8~kpc.

\subsection{Superburst phenomenology}

\begin{figure}[!t]
\includegraphics[width=\columnwidth]{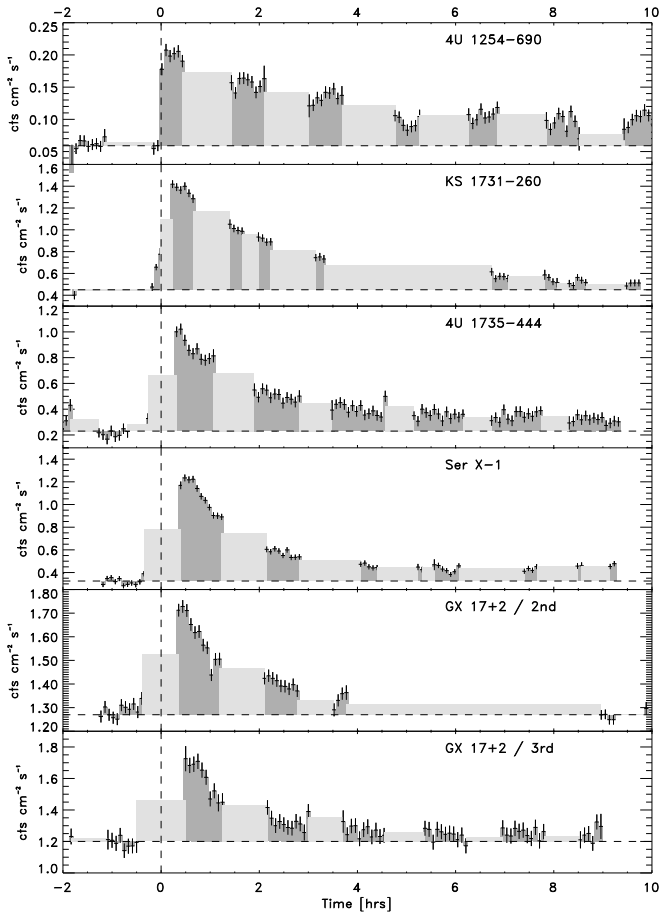}
\caption{2-28 keV light curves of 2 superbursts from \bron\ compared
with those of the other superbursts detected with the WFCs. The
e-folding decay times are from top to bottom 6.0 hr (4U~1254-69;
In~'t~Zand et al.  2003a), 2.7 hr (KS~1731-260; Kuulkers et al. 2002a),
1.4 hr (4U~1735-444; Cornelisse et al. 2000), 1.2~hr (Ser X-1;
Cornelisse et al.  2002), 1.9~hr (\bron, 2nd superburst) and 1.0~hr
(\bron, 3rd superburst). The horizontal dashed lines indicate the
pre-burst flux level, the vertical dashed lines the best estimate for
the burst onsets, and the light grey regions represent flux estimates
during data gaps being averages of the last data point before the data
gap and the first after.
\label{figlcall5}}
\end{figure}

The observations of decay rates provide a mixed view. The e-folding
decay times of 0.7-2.2~hr are not really faster than for the other
superbursters which range between 1 and 6 hr (see
Fig.~\ref{figlcall5}), but the scatter in both measurements is
large. If one compares the average decay times, a marginal difference
does emerge. The weight-averaged e-folding decay time for \bron\ is
$1.28\pm0.07$~hr.  If we discount the relatively short decay times for
4U~1820-303 and 4U~1636-536, the weight-averaged decay time for the
other superbursters is $1.89\pm0.05$~hr. This is 50\% larger.  The two
excluded systems may indeed be extraordinary: the exceptional status
of 4U~1820-303 has already been eluded, and 4U~1636-536 is the only
superburster for which multiple superbursts were detected (Kuulkers et
al. 2004).

The net observed unabsorbed peak flux is similar over all 4 cases at
$(0.8-1.5)\times10^{-8}$~\ecs (bolometric). These fluxes are not equal
to the Eddington limit for \bron, but at least 20\% less. They are 50
to 100\% of the peak flux measured for ordinary Eddington-limited
X-ray bursts. This is also observed in other superbursters except
4U~1820-303 and may be related to the gravitational redshift
being higher during superbursts because of lack of photospheric radius
expansion, in contrast to ordinary bursts. Alternatively, it is also
not unusual that ordinary bursts from one source show a range in peak
flux of tens of percents (even when considering only radius-expansion
bursts; e.g., In 't Zand et al. 2003b).

The bolometric fluence is measured to be between 4.0 and
$5.5\times10^{-5}$~erg~cm$^{-2}$. An exception may be the last
superburst which could be roughly twice as fluent. For an 8-12~kpc
distance this translates to $(3-10)\times10^{41}$~erg. This is between
50 and 100\% of the superbursts seen from other systems whose energy
output is between 6.5 and $10\times10^{41}$~erg (excluding again the
exceptional case of 4U~1820-303).

The superbursts were searched and found in data stretches when \bron\
is not in an obvious flaring state. The effective observing time on
\bron\ in this non-flaring state for superbursts is 10.2 Msec. Thus,
the mean recurrence time is $30\pm15$~d. The shortest actual observed
recurrence time is 8.2~d. The mean value is 15 times smaller than the
average value for all superbursters with persistent luminosities close
to 0.1 times Eddington (In 't Zand et al.  2004b).

In a number of superbursts reported in the literature, there is
evidence for quenching of ordinary bursting behavior for periods after
the superbursts between 1 week (in 4U 1735-44; Cornelisse et al. 2000)
and 1 month (in KS 1731-260, Kuulkers et al. 2002a, and Ser X-1,
Cornelisse et al. 2002). For \bron, regular bursts (see
Table~\ref{tab0}) were detected 2.2~d after flare 2, 10.5~d after
flare 3, 2.3~d after flare 4 and 12.8 days after flare number 6 (the
delays for flares 3 and 4 were determined from bursts detected with
RXTE and published in Kuulkers et al. 2002b). Therefore, quenching in
\bron, if at all present, is rather brief.

With regards to ignition, it is important to note that also for the
superbursts in \bron\ ordinary 'precursor' X-ray bursts were
marginally detected whenever possible. This appears to be a general
characteristic of any superburst. The question is whether the
precursor is what triggers the superburst (and is truly a precursor)
or vice versa. Given that precursors are often, but not always (e.g.,
4U 1636-536; see Kuulkers 2004), less energetic than other X-ray
bursts in the same object suggests that the precursor is merely a
result of the superburst and not responsible for the superburst
ignition. The precursors in \bron\ appear to be of the short
variety. The statistical quality of the data does not allow an
accurate check of whether these precursors are less energetic than
such bursts seen otherwise in \bron.

\subsection{Comparison with theoretical models}
\label{s53}

We now compare the observed properties of the superbursts with
theoretical models. We consider a 1.4 $M_\odot$ neutron star with
$R=10\ \mathrm{km}$, giving a surface gravity
$g=(GM/R^2)(1+z)=2.45\times 10^{14}\ \mathrm{cm\ s^{-2}}$ and redshift
$z=0.31$. The Eddington flux at the surface of the star is $F_{\rm
Edd}=cg/\kappa$, where $\kappa=0.2(1+X)\ {\rm cm^2\ g^{-1}}$ is the
electron scattering opacity, and $X$ is the surface hydrogen mass
fraction. The corresponding local Eddington accretion rate is given by
$F_{\rm Edd}=\dot m_{\rm Edd}c^2z/(1+z)$, where $c^2z/(1+z)\approx
GM/R$ is the gravitational energy release. For solar composition
$X=0.7$, we find $\dot m_{\rm Edd}=(c/\kappa R)(1+z/2)=1.02\times
10^5\ \mathrm{g\ cm^{-2}\ s^{-1}}$, which we take as our fiducial
accretion rate.

We first consider the overall energetics, using the mean properties of
the flares. The total radiated energy is $E_{41}=E_{\rm rad}/10^{41}\
\mathrm{ergs}=3.8\ (E_b/5\times 10^{-5}\ \mathrm{erg\ cm^{-2}})(d/8\
\mathrm{kpc})^2$, where $E_b$ is the observed fluence. The column
depth accreted between bursts is $y=\dot m \Delta t/(1+z)$, where
$\Delta t$ is the time measured by the observer, giving
$y_{11}=y/10^{11}\ \mathrm{g\ cm^{-2}}=2.0\ (t/30\ \mathrm{days})(\dot
m/\dot m_\mathrm{Edd})$. The implied energy release per gram is
$E_{\rm nuc}=E_{\rm rad}(1+z)/4\pi R^2 y$, giving
\begin{eqnarray}
E_{17} & = & 2.0\ \left({E_b\over 5\times 10^{-5}\ \mathrm{erg\ cm^{-2}}}\right)
\left({t\over 30\ \mathrm{days}}\right)^{-1} \nonumber \\
 & & \times
\left({R/d\over 10\ \mathrm{km\ @}\ 8\ \mathrm{kpc}}\right)^{-2}
\left({\dot m\over \dot m_\mathrm{Edd}}\right)^{-1}
\end{eqnarray}
where $E_{17}=E_{\rm nuc}/10^{17}\ \mathrm{erg\ g^{-1}}$. Since carbon
burning to iron releases $\approx 10^{18}\ \mathrm{erg\ g^{-1}}$, the
implied carbon fraction is $\approx 20$\%, although it may be a factor
of $\approx 2$ smaller if photodisintegration of heavy elements
enhances the energy release (Schatz et al.~2003a).

The inferred ignition column of $y_{11}\approx 2$ agrees very well
with the predictions of CB01. In Table~\ref{tab2}, we show ignition
conditions for accretion at $\dot m\approx \dot m_\mathrm{Edd}$
calculated following CB01. We assume that the layer is heated from
below by a flux $\dot mQ_b$ coming from the crust, and adopt the value
$Q_b=0.1$ MeV per nucleon found by Brown (2000). Schatz et al.~(1999)
calculated the products of stable hydrogen/helium burning at the
Eddington rate, and found that the rp-process produced mainly nuclei
with masses $56$--$68$. Here, we consider a mixture of carbon (mass
fraction $X_C$) and $^{64}$Ni (mass fraction $1-X_C$), and calculate
the resulting ignition conditions. The agreement with the observed
values is remarkable. For accretion at $\dot m=1.1\ \dot m_{\rm Edd}$, we find $y_{11}\approx 2$, in excellent agreement with the mean properties of the flares from \bron.

The inferred carbon mass fraction $X_C\approx 20$\% is somewhat higher
than found in calculations of rp-process hydrogen/helium
burning. Schatz et al.~(1999,2003b) find $X_C\approx 4$\% for stable
burning at $\dot m=\dot m_{\rm Edd}$. If the heavy nuclei are in the
$A\approx 100$ mass range rather than $A=64$--$68$,
photodisintegration reactions could enhance the energy release by a
factor of 2, giving $X_C\approx 10$\%. However, Schatz et al.~(2003b)
show that there is an inverse correlation between $X_C$ and $A$,
making $A\approx 64$--$68$ more likely for larger $X_C$. There are
significant uncertainties in some of the nuclear data required for
rp-process calculations (e.g.~Woosley et al.~2004); however, the high
$X_C$ inferred from the observations may prove problematic. Flare 4 is
particularly restrictive, having a recurrence time of only $8.2$ days,
giving
$E_{17}=4.4$. Such a short recurrence time can be obtained by an
increase of accretion rate by $\approx 50$\% (see Table~\ref{tab2});
however, the large burst energy requires $X_C=0.4$ if the energy
release is supplied entirely by carbon burning, a factor of two larger
than the mean value.

\renewcommand{\arraystretch}{1.5}
\renewcommand{\tabcolsep}{0.3cm}

\begin{table}
\caption{Ignition conditions for superbursts at near-Eddington accretion rates$^\ast$\label{tab2}}
\begin{center}
\begin{tabular}{ccccc}
\hline\hline
$\dot m$ ($\dot m_\mathrm{Edd}$) & $X_C$ & $y_{11}$ & $E_{41}$ &
$t_\mathrm{rec}\ (\mathrm{d})$ \\
\hline
1.1 & 0.2 & 2.4 & 4.5 & 31 \\  
1.8 & 0.4 & 0.97 & 3.7 & 8.0 \\
\hline\hline
\end{tabular}
\end{center}

\noindent
$^\ast$following CB01. We assume a flux from the crust of $Q_b=0.1$
MeV per nucleon. The local Eddington accretion rate is $\dot
m_\mathrm{Edd}=1.02\times 10^5\ \mathrm{g\ cm^{-2}\ s^{-1}}$. We take
the heavy nucleus to be $^{64}$Ni. A redshift correction of $1+z=1.31$
has been adopted to translate the superburst energy and recurrence time
into the observer's frame.
\end{table}

We have already noted that the exponential decay times of the flares from
GX17+2 are slightly shorter than those of other superbursts. Cumming \&
Macbeth (2004) studied the thermal evolution of the burning layers as
they cool following the thermal runaway. They showed that after a time
$0.7\ \mathrm{h}\ (y_{11}/2)^{3/4}(E_{17}/2)^{-1.1}(g_{14}/2.45)^{-5/4}$
(we do not show factors describing the composition of the layer), the
flux decays as a power law $\propto t^{-4/3}$. From their results, we
find that
\begin{equation}\label{eq:decay}
{F\over F_{\rm Edd}}\approx 0.14\ \left({t\over 1\
\mathrm{h}}\right)^{-4/3} \left({E_{17}\over
2}\right)^{1/2}
\left({y_{11}\over 2}\right)\left({g_{14}\over 2.45}\right)^{-5/3}
\end{equation}
during the power law decay, where we apply redshift corrections to the
flux and time. The observed decays are slower than predicted by
equation (\ref{eq:decay}). One way to see this is that after 2 hours,
we expect $F/F_\mathrm{Edd}\approx 6$\%, about a factor of two lower
than the observed value\footnote{We note that the dynamic range of the
WFC measurements is too small to be able to measure the power law
decay}.

CM04 also estimated the time for which normal Type I bursts would be
quenched following the superburst. They found that the critical flux
above which Type I X-ray bursts are stabilized is
$F_\mathrm{crit}\approx 5\times10^{22}\ (\dot m/\dot m_\mathrm{Edd})$
\ecs. Equation (\ref{eq:decay}) then gives a quenching
timescale $t_\mathrm{quench}\approx 23\ \mathrm{h}\
(y_{11}/2)^{3/4}(E_{17}/2)^{3/8}(\dot m/\dot
m_\mathrm{Edd})^{-3/4}$. For flare 4, the quenching timescale is
particularly short: taking $\dot m=1.5\ \dot m_\mathrm{Edd}$ and
$E_{17}=4$ gives $t_\mathrm{quench}=13$ hours. This is in good
agreement with the upper limits to the quenching times discussed
earlier.

\section{Conclusion}

In 10.2 Ms of effective observing time on GX17+2, we have identified 4
flares which are very likely superbursts. This is the first time that
superbursts are reported from a neutron star that is accreting matter near
the Eddington limit. Their properties are consistent with the smaller
ignition mass predicted by CB01 for these accretion rates. The average
recurrence time of 30 d is 15 times shorter than for superbursters that
accrete at 0.1 to 0.3 times the Eddington limit. The quenching time is
less than 2 days, also an order of magnitude shorter than low luminosity
superbursters. The decay rate and the radiation energy output of the
superbursts is approximately half that of the low luminosity
superbursters (excluding 4U~1820-303 and 4U~1636-56).

The observed recurrence times and quenching times agree very well with
the results of CB01 for carbon shell flashes at $\dot m\approx\dot
m_\mathrm{Edd}$. The inferred energy release of $E_{17}=2$--$4$ implies a
carbon fraction of $\approx 20$\%. This is larger than current
calculations of the products of H/He burning suggest (Schatz et al.~2003b;
Woosley et al.~2004), providing a new constraint on these models. In
addition, the decay timescales, although shorter than for superbursts at
lower accretion rates, are longer than the cooling models of CM04. A more
detailed comparison of the CM04 and observed lightcurves is in
progress; this should provide interesting constraints given the good
agreement of ignition models and the ignition mass inferred from the
energetics.

\acknowledgement 

We thank the referee, Lev Titarchuk, for a careful review and Erik
Kuulkers for very useful discussions and comments on an earlier
version of this paper. John Heise, Pietro Ubertini and Frank Verbunt
are acknowledged for proposing the Galactic Center observation
campaigns of the WFCs. Gerrit Wiersma, Jaap Schuurmans, Nuovo
Telespazio and the BeppoSAX Science Data Center are thanked for
continued support.  JZ acknowledges support from the Netherlands
Organization for Scientific Research (NWO).  AC is supported by NASA
through Hubble Fellowship grant HF-01138 awarded by the Space
Telescope Science Institute, which is operated by the Association of
Universities for Research in Astronomy, Inc., for NASA, under contract
NAS 5-26555.

\end{document}